\pdfoutput=1

\documentclass[11pt,twoside,a4paper,cmspaper,final,collab]{cms-tdr}
\def\svnVersion{289260}\def\cmsCernNo{2013-037}\def\cmsCernDate{\today}\def\cmsMessage{Published in Physical Review Letters as \href{http://dx.doi.org/10.1103/PhysRevLett.114.191802}{\doi{10.1103/PhysRevLett.114.191802.}}}

\begin{document}\cmsNoteHeader{BPH-14-001}

\hyphenation{had-ron-i-za-tion}
\hyphenation{cal-or-i-me-ter}
\hyphenation{de-vices}
\RCS$HeadURL: svn+ssh://svn.cern.ch/reps/tdr2/papers/BPH-14-001/trunk/BPH-14-001.tex $
\RCS$Id: BPH-14-001.tex 282621 2015-03-28 21:21:00Z alverson $
\newlength\cmsFigWidth
\ifthenelse{\boolean{cms@external}}{\setlength\cmsFigWidth{0.85\columnwidth}}{\setlength\cmsFigWidth{0.4\textwidth}}
\ifthenelse{\boolean{cms@external}}{\providecommand{\cmsLeft}{top}}{\providecommand{\cmsLeft}{left}}
\ifthenelse{\boolean{cms@external}}{\providecommand{\cmsRight}{bottom}}{\providecommand{\cmsRight}{right}}
\providecommand{\PgUn}{\ensuremath{\PgU\mathrm{(}n\mathrm{S)}}\xspace}
\ifthenelse{\boolean{cms@external}}{\providecommand{\suppMaterial}{the supplemental material~\cite{bib:suppMaterial}\xspace}} {\providecommand{\suppMaterial}{Appendix~\ref{app:tables}\xspace}}
\ifthenelse{\boolean{cms@external}}{\providecommand{\suppMaterialSecond}{the supplemental material\xspace}} {\providecommand{\suppMaterialSecond}{Appendix~\ref{app:tables}\xspace}}
\ifthenelse{\boolean{cms@external}}{\providecommand{\suppMaterialRef}[2]{#1}}{\providecommand{\suppMaterialRef}[2]{#2}}

\newcommand{\QQbar}{\ensuremath{Q \overline{Q}}\xspace}
\newcommand{\PsiOne}{\PJGy}
\newcommand{\PsiTwo}{\Pgy\xspace}
\newcommand{\tnp}{T\&P}
\newcommand*\red{\color{red}}

\cmsNoteHeader{BPH-14-001}
\title{\texorpdfstring{Measurement of prompt \PsiOne and \PsiTwo double-differential cross sections in $\Pp\Pp$ collisions at $\sqrt{s} = 7$\TeV}{Measurement of J/psi and psi(2S) prompt double-differential cross sections in pp collisions at sqrt(s) = 7 TeV}}

\date{\today}

\abstract{The double-differential cross sections of promptly produced \PsiOne and \PsiTwo mesons
are measured in $\Pp\Pp$ collisions at $\sqrt{s} = 7$\TeV,
as a function of transverse momentum \pt and absolute rapidity $\abs{y}$.
The analysis uses \PsiOne and \PsiTwo dimuon samples collected by 
the CMS experiment,
corresponding to integrated luminosities of 4.55 and 4.90\fbinv, respectively.
The results are based on a two-dimensional analysis of the dimuon invariant mass and decay length,
and extend to $\pt=120$ and 100\GeV for the \PsiOne and \PsiTwo, respectively,
when integrated over the interval $\abs{y}<1.2$.
The ratio of the \PsiTwo to \PsiOne cross sections is also reported for $\abs{y}<1.2$,
over the range $10 < \pt < 100$\GeV.
These are the highest \pt values for which the cross sections and ratio have been measured.}

\hypersetup{%
pdfauthor={CMS Collaboration},%
pdftitle={Measurement of J/psi and psi(2S) prompt double-differential cross sections in pp collisions at sqrt(s) = 7 TeV},%
pdfsubject={CMS},%
pdfkeywords={CMS, charmonium, cross sections}}

\maketitle

Studies of heavy-quarkonium production are of central importance for an improved understanding of
nonperturbative quantum chromodynamics (QCD)~\cite{bib:Brambilla2011}.
The nonrelativistic QCD (NRQCD) effective-field-theory framework~\cite{bib:NRQCD}, 
arguably the best formalism at this time,
factorizes high-\pt quarkonium production in short-distance and long-distance scales.
First a heavy quark-antiquark pair, \QQbar, is produced in a Fock state $^{2S+1}L_J^{[a]}$, 
with spin $S$, orbital angular momentum $L$, and total angular momentum $J$ 
that are either identical to (color singlet, $a=1$) or different from (color octet, $a=8$) 
those of the corresponding quarkonium state.
The \QQbar cross sections are determined by short-distance coefficients (SDC),
kinematic-dependent functions calculable perturbatively as expansions in
the strong-coupling constant $\alpha_s$.
Then this ``preresonant" \QQbar pair binds into the physically observable quarkonium
through a nonperturbative evolution that may change $L$ and $S$,
with bound-state formation probabilities proportional to long-distance matrix elements (LDME).
The LDMEs are conjectured to be constant (i.e., independent of the \QQbar momentum) 
and universal (i.e., process independent).
The color-octet terms are expected to scale with powers of the heavy-quark velocity in the \QQbar rest frame.
In the nonrelativistic limit, an \mbox{$S$-wave} vector quarkonium state should be formed from a \QQbar pair produced
as a color singlet ($^3S_1^{[1]}$) or as one of three color octets ($^1S_0^{[8]}$, $^3S_1^{[8]}$, and $^3P_J^{[8]}$).

Three ``global fits'' to measured quarkonium data~\cite{bib:BK, bib:GWWZ, bib:Chao}
obtained incompatible octet LDMEs, despite the use of essentially identical theory 
inputs: next-to-leading-order (NLO) QCD calculations of the singlet and octet SDCs.
The disagreement stems from the fact that different sets of measurements were considered.
In particular, the results crucially depend on the minimum \pt of the fitted measurements~\cite{bib:FKLSW},
because the octet SDCs have different \pt dependences.
Fits including low-\pt cross sections lead to the conclusion that, at high \pt,
quarkonium production should be dominated by transversely polarized octet terms.
This prediction is in stark contradiction with the unpolarized production seen by the
CDF~\cite{bib:CDF-psi-pol,bib:CDF-upsilon-pol}
and CMS~\cite{bib:CMS-upsilon-pol,bib:CMS-psi-pol} experiments,
an observation known as the ``quarkonium polarization puzzle".
As shown in Ref.~\cite{bib:FKLSW}, the puzzle is seemingly solved by restricting the NRQCD 
global fits to high-\pt quarkonia, indicating that the presently
available fixed-order calculations provide SDCs unable to reproduce reality at lower \pt values
or that NRQCD factorization only holds for \pt values much larger than the quarkonium mass.
The polarization measurements add a crucial dimension to the global fits because 
the various channels have remarkably distinct polarization properties: 
in the helicity frame, $^3S_1^{[1]}$ is longitudinally polarized, 
$^1S_0^{[8]}$ is unpolarized, $^3S_1^{[8]}$ is transversely polarized, and 
$^3P_J^{[8]}$ has a polarization that changes significantly with \pt.
Bottomonium and prompt charmonium polarizations reaching or exceeding $\pt = 50$\GeV 
were measured by CMS~\cite{bib:CMS-upsilon-pol,bib:CMS-psi-pol}, 
using a very robust analysis framework~\cite{bib:LTGen, bib:ImprovedQQbarPol},
on the basis of event samples collected in 2011.
Instead, the differential charmonium cross sections published by CMS~\cite{bib:BPH-10-014}
are based on data collected in 2010 and have a much lower \pt reach.
Measurements of prompt charmonium cross sections extending well beyond 
$\pt = 50$\GeV will trigger improved NRQCD global fits, restricted to a kinematic
domain where the factorization formalism is unquestioned, and will provide more 
accurate and reliable LDMEs.

This Letter presents measurements of the double-differential cross sections of \PsiOne and \PsiTwo mesons 
promptly produced in \Pp\Pp\ collisions at a center-of-mass energy of 7\TeV,
based on dimuon event samples collected by CMS in 2011.
They complement other prompt charmonium cross sections measured at the 
LHC, by ATLAS~\cite{bib:Aad:2011sp,bib:Aad:2014fpa}, LHCb~\cite{bib:LHCb1S,bib:LHCb2S}, and ALICE~\cite{bib:ALICE}.
The analysis is
made in four
bins of absolute rapidity ($\abs{y}<0.3$, $0.3<\abs{y}<0.6$, $0.6<\abs{y}<0.9$, and $0.9<\abs{y}<1.2$)
and in the \pt ranges 10--95\GeV for the \PsiOne and 10--75\GeV for the \PsiTwo.
A rapidity-integrated result in the range $\abs{y}<1.2$ is also provided, extending the \pt reach to
120\GeV for the \PsiOne and 100\GeV for the \PsiTwo.
The corresponding \PsiTwo over \PsiOne cross section ratios are also reported.
The dimuon invariant mass distribution is used to separate the \PsiOne and \PsiTwo signals from other processes,
mostly pairs of uncorrelated muons, while the dimuon decay length is used to separate the nonprompt charmonia,
coming from decays of \PQb hadrons, from the prompt component.
Feed-down from decays of heavier charmonium states, approximately 33\% of the prompt \PsiOne
cross section~\cite{bib:Faccioli-feeddown}, is not distinguished from the directly produced charmonia.

The CMS apparatus is based on
a superconducting solenoid of 6\unit{m} internal diameter, providing a 3.8\unit{T} field.
Within the solenoid volume are a silicon pixel and strip tracker, a lead tungstate crystal
electromagnetic calorimeter, and a brass and scintillator hadron calorimeter. Muons are measured with
three kinds of
gas-ionization detectors:
drift tubes, cathode strip chambers, and resistive-plate chambers.
The main subdetectors used in this analysis are the silicon tracker and the muon system, which
enable the measurement of muon momenta over the pseudorapidity range $\abs{\eta} < 2.4$.
A more detailed description of the CMS detector, together with a definition of the coordinate system
used and the relevant kinematic variables, can be found in Ref.~\cite{Chatrchyan:2008zzk}.

The events were collected using a two-level trigger system.
The first level, made of custom hardware processors, uses data from the muon system to select events
with two muon candidates. The high-level trigger, adding information from the silicon tracker,
reduces the rate of stored events by requiring an opposite-sign muon pair of invariant mass
$2.8<M<3.35$\GeV, $\pt > 9.9$\GeV, and $\abs{y} < 1.25$ for the \PsiOne trigger,
and $3.35 <M < 4.05$\GeV and $\pt > 6.9$\GeV for the \PsiTwo trigger.
No \pt requirement is imposed on the single muons at trigger level.
Both triggers require a dimuon vertex fit $\chi^2$ probability greater than 0.5\% and
a distance of closest approach between the two muons
less than 5\unit{mm}.
Events where the muons bend towards each other in the magnetic field are rejected to lower the trigger rate
while retaining the highest-quality dimuons.
The \PsiOne and \PsiTwo analyses are conducted independently, using event samples separated at the trigger level.
The \PsiTwo sample corresponds to an integrated luminosity of 4.90\fbinv,
while the \PsiOne sample has a reduced value, 4.55\fbinv, because the
\pt threshold of the \PsiOne trigger was raised to 12.9\GeV in a fraction of the data-taking period;
the integrated luminosities have an uncertainty of 2.2\%~\cite{bib:lumi}.

The muon tracks are required to have hits in at least eleven tracker layers, with at least two in the
silicon pixel detector, and to be matched with at least one segment in the muon system.
They must have a good track fit quality ($\chi^2$ per degree of freedom smaller than 1.8)
and point to the interaction region.
The selected muons must also match in pseudorapidity and azimuthal angle with
the muon objects responsible for triggering the event.
The analysis is restricted to muons produced within a fiducial phase-space window where the muon detection efficiencies
are accurately measured: $\pt > 4.5$, 3.5, and 3.0\GeV for the regions
$\abs{\eta} < 1.2$, $1.2 < \abs{\eta} < 1.4$, and $1.4 < \abs{\eta} < 1.6$, respectively.
The combinatorial dimuon background is reduced by requiring a dimuon vertex fit $\chi^2$ probability larger than 1\%.
After applying the event selection criteria,
the combined yields of prompt and nonprompt charmonia in the range $\abs{y}<1.2$
are 5.45\,M for the \PsiOne and 266\,k for the \PsiTwo.
The prompt charmonia are separated from those resulting from decays of  \PQb hadrons
through the use of the dimuon pseudo-proper decay length~\cite{bib:BPH-10-002},
$\ell = L_{xy} \, M / \pt$, 
where $L_{xy}$ is the transverse decay length in the laboratory frame, measured after removing the two muon tracks
from the calculation of the primary vertex position. For events with multiple collision vertices,
$L_{xy}$ is calculated with respect to the vertex closest to the direction of the dimuon momentum,
extrapolated towards the beam line.

For each $(\abs{y},\pt)$ bin, the prompt charmonium yields are evaluated through an
extended unbinned maximum-likelihood fit to the two-dimensional $(M,\ell)$ event distribution.
In the mass dimension, the shape of each signal peak is represented by a Crystal Ball (CB) function~\cite{bib:CrystalBall},
with free mean ($\mu_\mathrm{CB}$) and width ($\sigma_\mathrm{CB}$) parameters.
Given the strong correlation between the two CB tail parameters, $\alpha_\mathrm{CB}$ and $n_\mathrm{CB}$,
they are fixed to values evaluated from fits to event samples integrated in broader \pt ranges.
A single CB function provides a good description of the signal mass peaks,
given that the dimuon mass distributions are studied in narrow $(\abs{y},\pt)$ bins,
within which the dimuon invariant mass resolution has a negligible variation.
The mass distribution of the underlying continuum background is described by an exponential function.
Concerning the pseudo-proper decay length variable,
the prompt signal component is modeled by a resolution function,
which exploits the per-event uncertainty information provided by the vertex reconstruction algorithm,
while the nonprompt charmonium term is modeled by an exponential function convolved with the resolution function.
The continuum background component is represented by a sum of prompt and nonprompt empirical forms.
The distributions are well described with a relatively small number of free parameters.

\begin{figure*}[thb]
\centering
\includegraphics[width=0.48\textwidth]{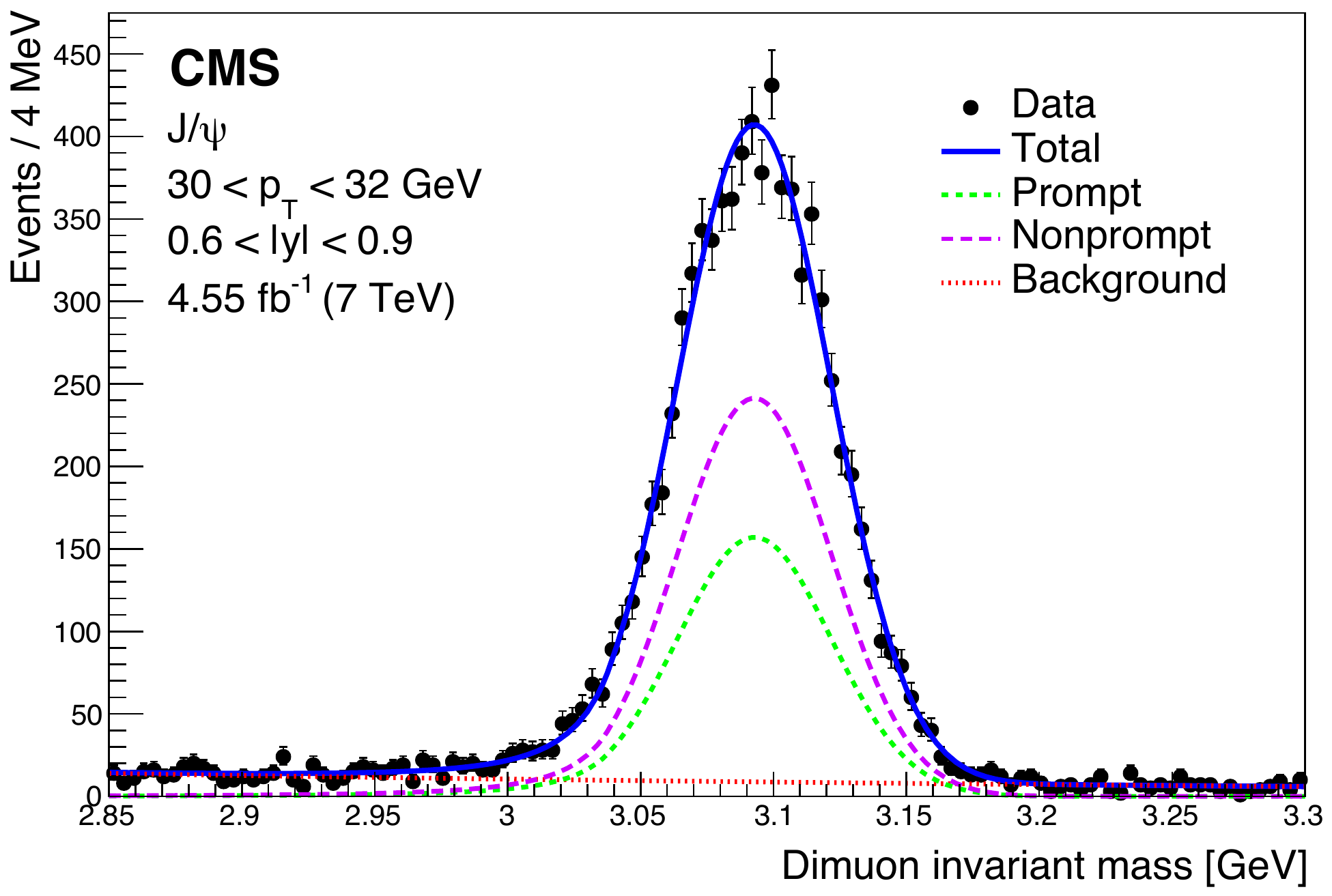}
\includegraphics[width=0.48\textwidth]{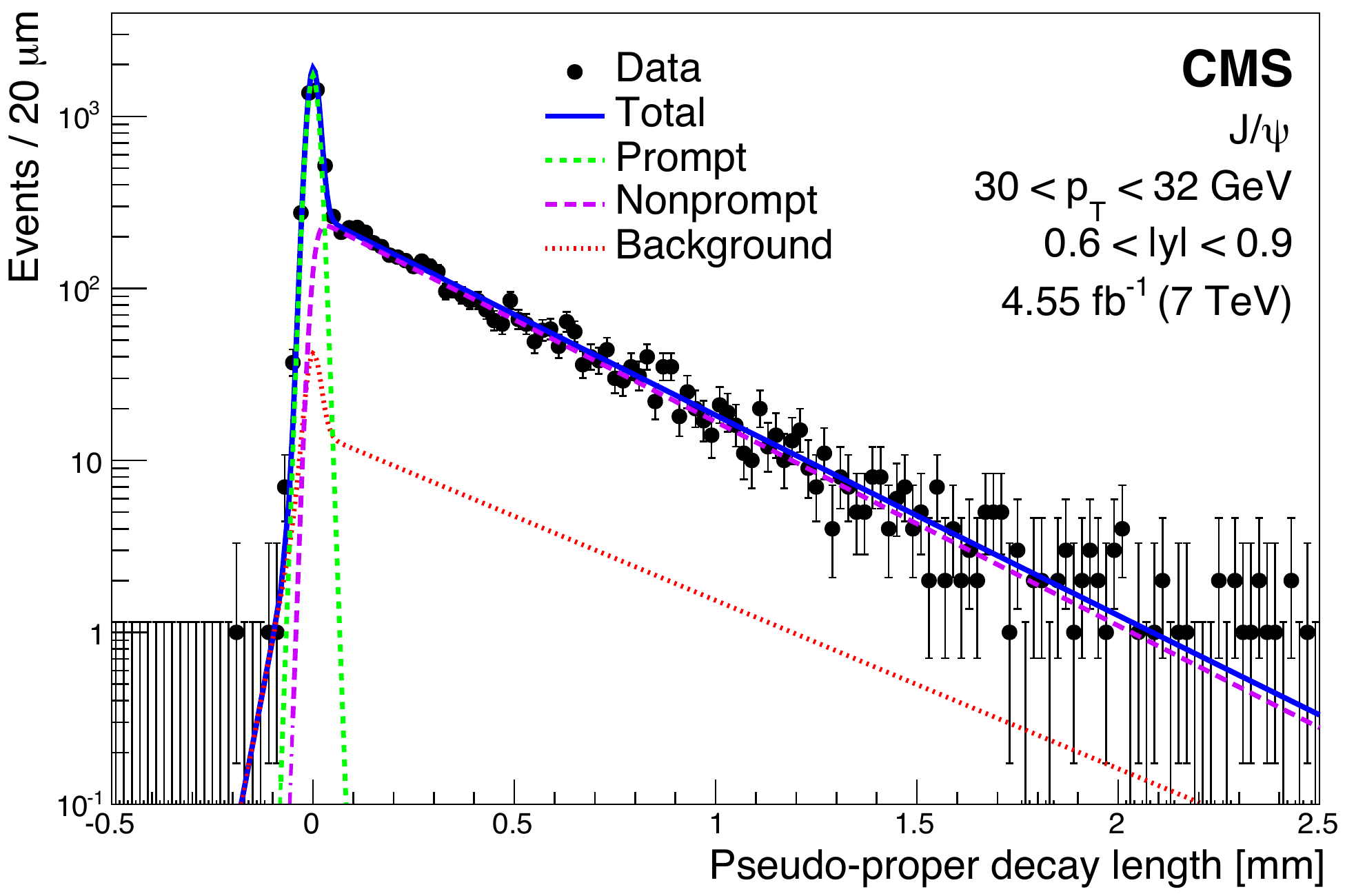}
\includegraphics[width=0.48\textwidth]{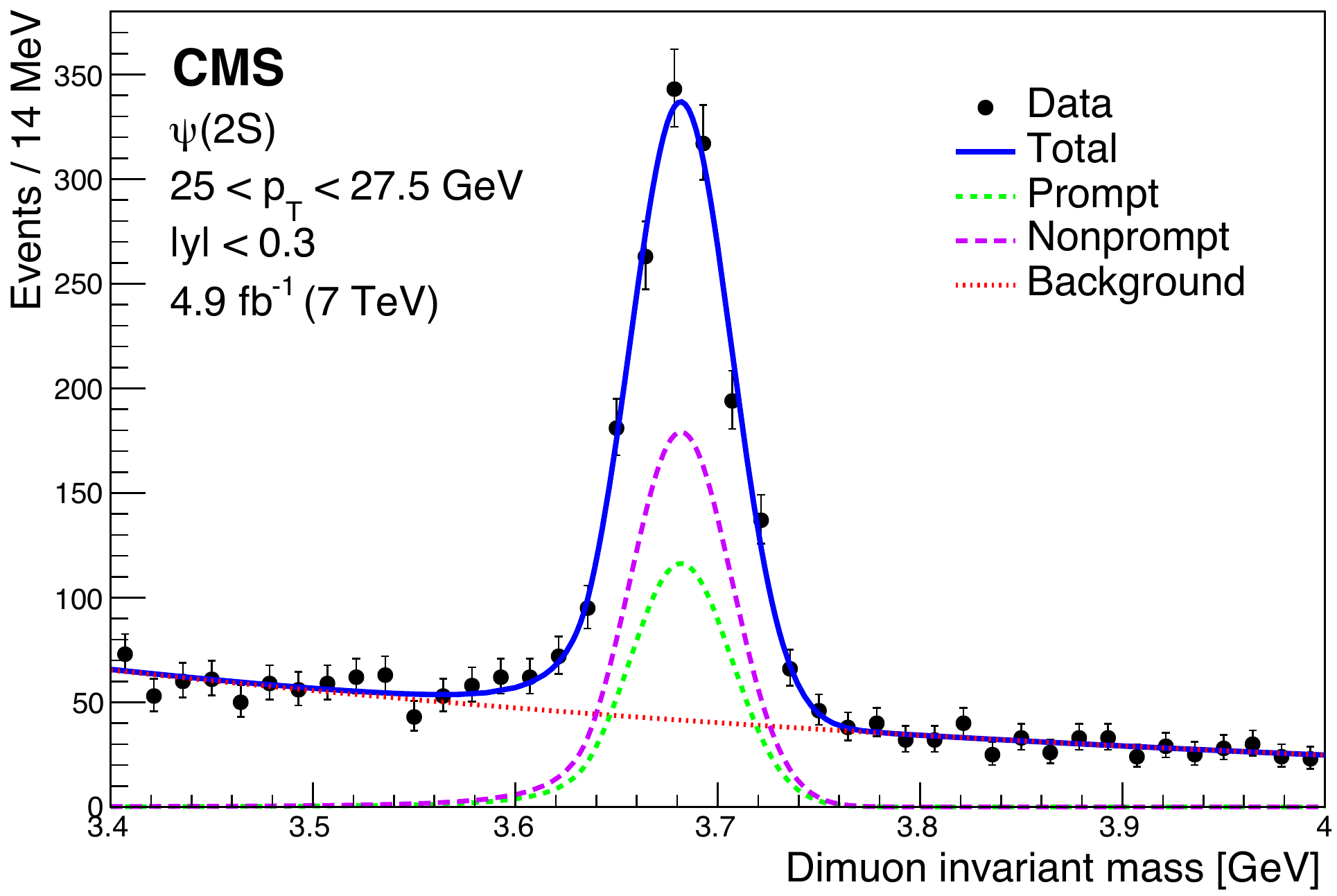}
\includegraphics[width=0.48\textwidth]{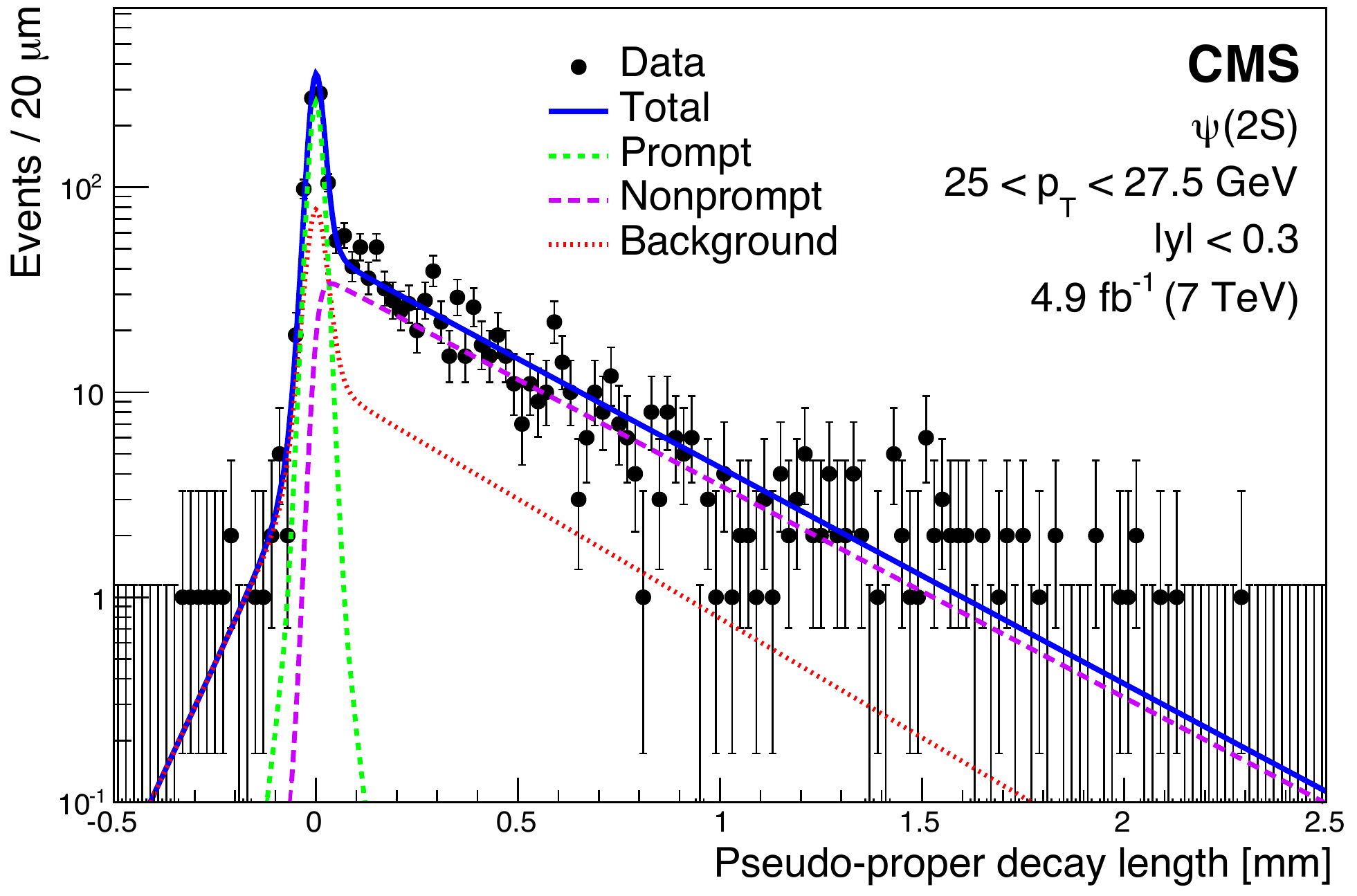}
\caption{Projections on the dimuon invariant mass (left) and pseudo-proper decay length (right) axes,
for the \PsiOne (top) and \PsiTwo (bottom) events in the kinematic bins given in the plots.
The right panels show dimuons of invariant mass within $\pm$3\,$\sigma_\mathrm{CB}$ of the pole masses.
The curves, identified in the legends, represent the result of the fits described in the text.
The vertical bars on the data points show the statistical uncertainties.}
\label{fig:fit-projections}
\end{figure*}

Figure~\ref{fig:fit-projections} shows the \PsiOne and \PsiTwo
dimuon invariant mass and pseudo-proper decay length projections for two representative $(\abs{y},\pt)$ bins.
The decay length projections are shown for events with dimuon invariant mass within ${\pm}3\,\sigma_\mathrm{CB}$ of the pole mass.
In the highest \pt bins, where the number of dimuons is relatively small, stable results are obtained by fixing
$\mu_\mathrm{CB}$ and the slope of the exponential-like function describing the nonprompt combinatorial background
to values extrapolated from the trend found from the lower-\pt bins.
The systematic uncertainties in the signal yields are evaluated by repeating the
fit with different functional forms,
varying the values of the fixed parameters, and allowing for more free parameters in the fit.
The fit results are robust with respect to changes in the procedure;
the corresponding systematic uncertainties are negligible at low \pt and increase to
$\approx$2\% for the \PsiOne and $\approx$6\% for the \PsiTwo in the highest \pt bins.

The single-muon detection efficiencies $\epsilon_\mu$ are measured with a ``tag-and-probe" (\tnp) technique~\cite{Khachatryan:2010xn},
using event samples collected with triggers specifically designed for this purpose, including a sample enriched in dimuons
from \PsiOne decays where a muon is combined with another track
and the pair is required to have an invariant mass within the range 2.8--3.4\GeV.
The procedure was validated in the phase-space window of the analysis with detailed Monte Carlo (MC) simulation studies.
The measured efficiencies are parametrized as a function of muon \pt, in eight bins of muon $\abs{\eta}$.
Their uncertainties, reflecting the statistical precision of the \tnp\ samples and possible imperfections of the parametrization,
are $\approx$2--3\%.
The efficiency of the dimuon vertex fit $\chi^2$ probability requirement is also measured with the \tnp\ approach,
using a sample of events collected with a dedicated (prescaled) trigger.
It is around 95--97\%, improving with increasing \pt, with a 2\% systematic uncertainty
At high \pt, when the two muons might be emitted relatively close to each other, the efficiency of the dimuon trigger $\epsilon_{\mu\mu}$
is smaller than the product of the two single-muon efficiencies~\cite{bib:BPH-10-014},
$\epsilon_{\mu\mu} = \epsilon_{\mu_1} \, \epsilon_{\mu_2} \, \rho$.
The correction factor $\rho$ is evaluated with MC simulations, validated from data collected with single-muon triggers.
For $\pt < 35$\GeV, $\rho$ is consistent with being unity, within a systematic uncertainty estimated as 2\%, 
except in the $0.9<\abs{y}<1.2$ bin, 
where the uncertainty increases to 4.3\% for the \PsiOne\ if $\pt<12$\GeV, and to 2.7\% for the \PsiTwo\ if $\pt<11$\GeV. 
For $\pt > 35$\GeV, $\rho$ decreases approximately linearly with \pt, reaching 60--70\% for $\pt \sim 85$\GeV, 
with systematic uncertainties evaluated by comparing the MC simulation results with estimations made using data 
collected with single-muon triggers: 5\% up to $\pt=50$~(55)\GeV for the \PsiOne\ (\PsiTwo) and 10\% for higher \pt.
The total dimuon detection efficiency increases from $\epsilon_{\mu\mu} \approx78\%$ at $\pt = 15$\GeV to $\approx$85\% at 30\GeV,
and then decreases to $\approx$65\% at 80\GeV.

To obtain the charmonium cross sections in each $(\abs{y},\pt)$ bin without any restrictions on the kinematic variables of the two muons,
we correct for the corresponding dimuon acceptance, defined as
the fraction of dimuon decays having both muons emitted within the single-muon fiducial phase space.
These acceptances are calculated using a detailed MC simulation of the CMS experiment.
Charmonia are generated using a flat rapidity distribution
and \pt distributions based on previous measurements~\cite{bib:BPH-10-014};
using flat \pt distributions leads to negligible changes.
The particles are decayed by \EVTGEN~\cite{bib:EvtGen} interfaced to \PYTHIA~6.4~\cite{bib:Pythia},
while \PHOTOS~\cite{bib:PHOTOS2} is used to simulate final-state radiation.
The fractions of \PsiOne and \PsiTwo dimuon events in a given $(\abs{y},\pt)$ bin with both muons surviving the fiducial selections
depend on the decay kinematics and, in particular, on the polarization of the mother particle.
Acceptances are calculated using polarization scenarios corresponding to different values of the polar anisotropy parameter
in the helicity frame, $\lambda_\vartheta^\mathrm{HX}$:
$0$ (unpolarized), $+1$ (transverse), and $-1$ (longitudinal).
A fourth scenario, corresponding to $\lambda_\vartheta^\mathrm{HX} = +0.10$ for the \PsiOne and $+0.03$ for the \PsiTwo,
reflects the results published by CMS~\cite{bib:CMS-psi-pol}.
The two other parameters characterizing the dimuon angular distributions~\cite{bib:EPJC69},
$\lambda_\varphi$ and $\lambda_{\vartheta\varphi}$,
have been measured to be essentially zero~\cite{bib:CMS-psi-pol}
and have a negligible influence on the acceptance.
The acceptances are essentially identical for the two charmonia and are almost rapidity independent for $\abs{y}<1.2$.
The two-dimensional acceptance maps are calculated with large MC simulation samples, so that statistical fluctuations are small,
and in narrow $\abs{y}$ bins, so that variations within the bins can be neglected.
Since the efficiencies and acceptances are evaluated for events where the two muons bend away from each other,
a factor of two is applied to obtain the final cross sections.

\begin{figure}[t!h]
\centering
\includegraphics[width=0.47\textwidth]{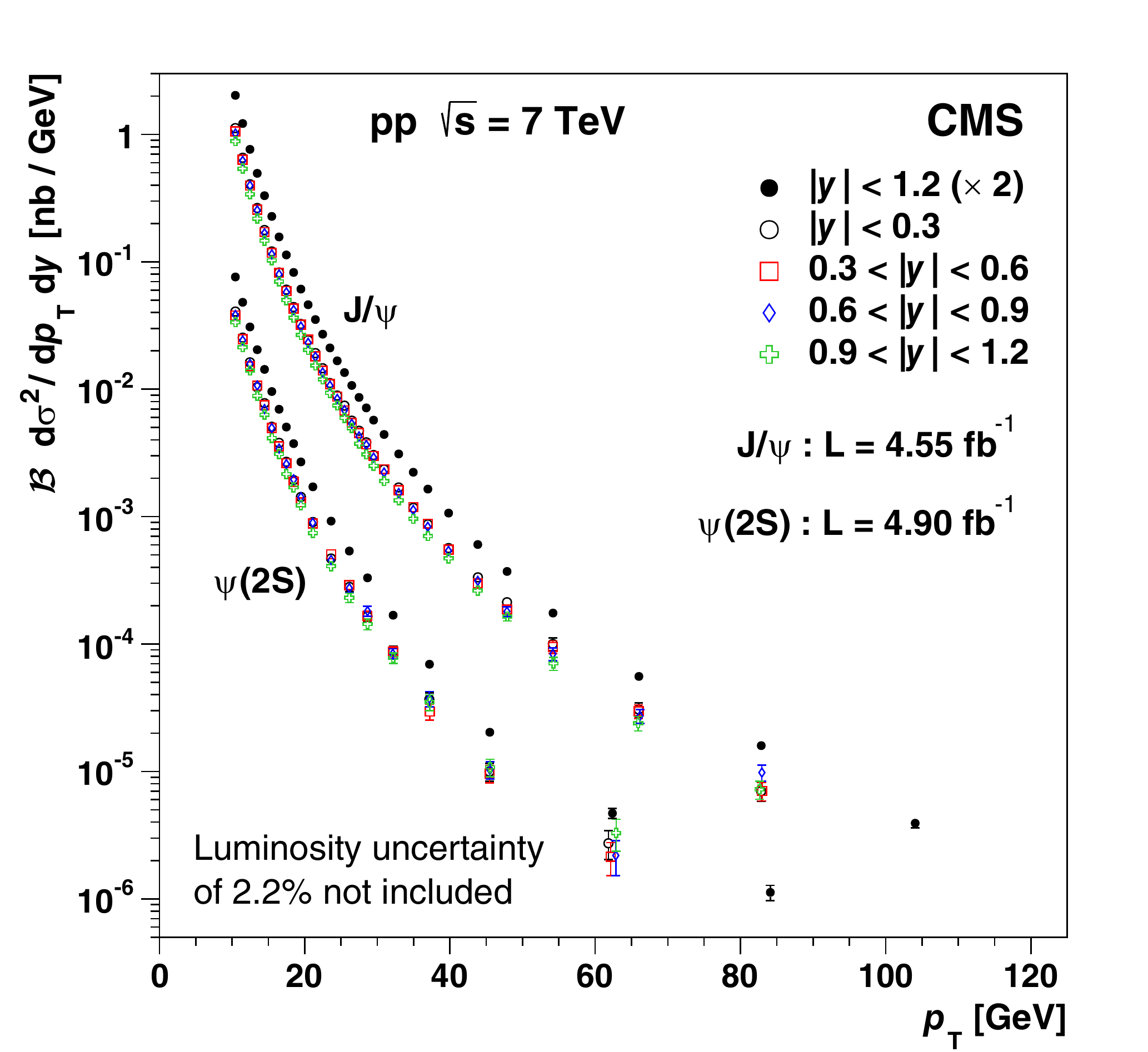}
\caption{The \PsiOne and \PsiTwo differential \pt cross sections times the dimuon branching fractions for four rapidity bins and
integrated over the range $\abs{y}<1.2$ (scaled up by a factor of 2 for presentation purposes), assuming the unpolarized scenario.
The vertical bars show the statistical and systematic uncertainties added in quadrature.}
\label{fig:Xsection}
\end{figure}

\begin{figure}[t!h]
\centering
\includegraphics[width=0.47\textwidth]{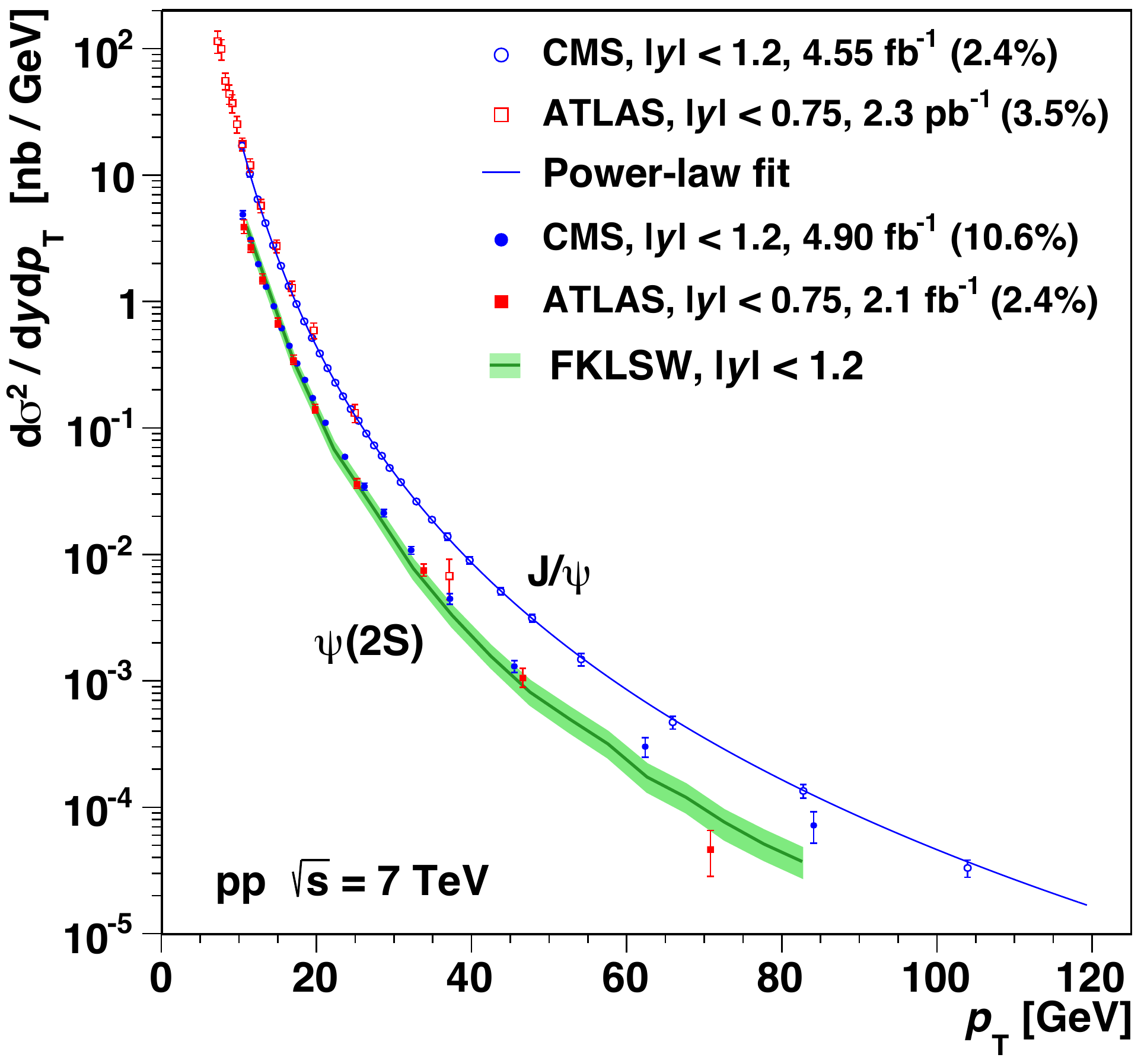}
\caption{The \PsiOne (open symbols) and \PsiTwo (closed symbols) differential (unpolarized) cross sections
from this analysis (circles) and from ATLAS (squares)~\cite{bib:Aad:2011sp,bib:Aad:2014fpa}.
The vertical bars show the statistical and systematic uncertainties added in quadrature, not including the
uncertainties from integrated luminosities and branching fractions, which are indicated by the percentages given in the legend.
The curve shows a fit of the \PsiOne cross section measured in this analysis to a power-law function,
while the band labelled FKLSW represents a calculation of the \PsiTwo cross section
using LDMEs determined with lower-\pt LHC data~\cite{bib:FKLSW}.}
\label{fig:XsectionTH}
\end{figure}

The double-differential cross sections of promptly produced \PsiOne and \PsiTwo in the dimuon channel,
$\mathcal{B}\, \rd^2\sigma/\rd\pt\, \rd{}y$,
where $\mathcal{B}\,$ is the \PsiOne or \PsiTwo dimuon branching fraction,
is obtained by dividing the fitted prompt-signal yields, already corrected on an event-by-event basis for
efficiencies and acceptance, by the integrated luminosity and the widths of the \pt and $\abs{y}$ bins.
The numerical values, including the relative statistical and systematic uncertainties,
are reported for both charmonia, five rapidity intervals, and four polarization scenarios
in Tables~\suppMaterialRef{1--4}{\ref{tab:xs-JPsi-rap}--\ref{tab:xs-PsiP-rap0}} of \suppMaterial.
Figure~\ref{fig:Xsection} shows the results obtained in the unpolarized scenario.
With respect to the $\abs{y}<0.3$ bin,
the cross sections drop by $\approx$5\% for $0.6<\abs{y}<0.9$ and $\approx$15\% for $0.9<\abs{y}<1.2$.
Measuring the charmonium production cross sections in the broader rapidity range $\abs{y}<1.2$ has the advantage that
the increased statistical accuracy allows the measurement to be extended to higher-\pt values,
where comparisons with theoretical calculations are particularly informative.
Figure~\ref{fig:XsectionTH} compares the rapidity-integrated (unpolarized) cross sections,
after rescaling with the branching fraction $\mathcal{B}$ of the dimuon decay channels~\cite{bib:PDG2014},
with results reported by ATLAS~\cite{bib:Aad:2011sp,bib:Aad:2014fpa}.
The curve represents a fit of the \PsiOne cross section measured in this analysis to a power-law function~\cite{bib:HERAb}.
The band labelled FKLSW represents 
the result of a global fit~\cite{bib:FKLSW} comparing SDCs calculated at NLO~\cite{bib:BK} with \PsiTwo
cross sections and polarizations previously reported by
CMS~\cite{bib:BPH-10-014,bib:CMS-psi-pol} and LHCb~\cite{bib:LHCb2S}.
According to that fit, \PsiTwo mesons are produced predominantly unpolarized.
At high \pt, the values reported in this Letter tend to be higher than the band, 
which is essentially determined from results for $\pt < 30$\GeV.

The ratio of the \PsiTwo to \PsiOne differential cross sections is also measured in the $\abs{y} < 1.2$ range,
recomputing the \PsiOne values in the \pt bins of the \PsiTwo analysis.
The measured values are reported in Table~\suppMaterialRef{5}{\ref{tab:ratio-xs}}
of \suppMaterial.
The corrections owing to the integrated luminosity, acceptances, and efficiencies
cancel to a large extent in the measurement of the ratio.
The total systematic uncertainty, dominated by the $\rho$ correction for $\pt > 30$\GeV
and by the acceptance and efficiency corrections for $\pt < 20$\GeV,
does not exceed 3\%, except for $\pt > 75$\GeV, where it reaches 5\%.
Larger event samples are needed to clarify the trend of the ratio for \pt above $\approx$35\GeV.

In summary, the double-differential cross sections of the \PsiOne and \PsiTwo mesons promptly produced in $\Pp\Pp$ collisions
at $\sqrt{s} = 7$\TeV have been 
measured
as a function of \pt in four $\abs{y}$ bins, 
as well as integrated over the $\abs{y}<1.2$ range, extending up to or beyond $\pt = 100$\GeV.
New global fits of cross sections and polarizations, including these high-\pt measurements,
will probe the theoretical calculations
in a kinematical region where NRQCD factorization is believed to be most reliable.
The new data should also provide input to stringent tests of recent theory developments, 
such as those described in Refs.~\cite{bib:Kang2011zza,bib:Kang2014tta,bib:Bodwin2014gia}.

\begin{acknowledgments}
We congratulate our colleagues in the CERN accelerator departments for the excellent performance of the LHC and thank the technical and administrative staffs at CERN and at other CMS institutes for their contributions to the success of the CMS effort. In addition, we gratefully acknowledge the computing centres and personnel of the Worldwide LHC Computing Grid for delivering so effectively the computing infrastructure essential to our analyses. Finally, we acknowledge the enduring support for the construction and operation of the LHC and the CMS detector provided by the following funding agencies: BMWFW and FWF (Austria); FNRS and FWO (Belgium); CNPq, CAPES, FAPERJ, and FAPESP (Brazil); MES (Bulgaria); CERN; CAS, MoST, and NSFC (China); COLCIENCIAS (Colombia); MSES and CSF (Croatia); RPF (Cyprus); MoER, ERC IUT and ERDF (Estonia); Academy of Finland, MEC, and HIP (Finland); CEA and CNRS/IN2P3 (France); BMBF, DFG, and HGF (Germany); GSRT (Greece); OTKA and NIH (Hungary); DAE and DST (India); IPM (Iran); SFI (Ireland); INFN (Italy); MSIP and NRF (Republic of Korea); LAS (Lithuania); MOE and UM (Malaysia); CINVESTAV, CONACYT, SEP, and UASLP-FAI (Mexico); MBIE (New Zealand); PAEC (Pakistan); MSHE and NSC (Poland); FCT (Portugal); JINR (Dubna); MON, RosAtom, RAS and RFBR (Russia); MESTD (Serbia); SEIDI and CPAN (Spain); Swiss Funding Agencies (Switzerland); MST (Taipei); ThEPCenter, IPST, STAR and NSTDA (Thailand); TUBITAK and TAEK (Turkey); NASU and SFFR (Ukraine); STFC (United Kingdom); DOE and NSF (USA). \end{acknowledgments}

\bibliography{auto_generated}
\ifthenelse{\boolean{cms@external}}{}{
    \clearpage
    \appendix
    \numberwithin{table}{section}
    \section{Tables of cross sections\label{app:tables}}
    
\begin{table*}[htbp]
\centering
\topcaption{The \PsiOne differential cross section times dimuon branching fraction
$\mathcal{B}\, \rd\sigma/\rd{}\pt$ in four rapidity ranges for the unpolarized scenario.
The relative uncertainties (first statistical and then systematic) are given in percent.
The systematic uncertainties are to be treated as bin-to-bin correlated.}
\label{tab:xs-JPsi-rap}
\begin{scotch}{*{5}{c}}
\rule{0pt}{0.4cm}$\Delta \pt$ & \multicolumn{4}{c}{$\mathcal{B}\, \rd\sigma/\rd{}\pt$ [pb/\GeVns{}]} \\ \cline{2-5}
\rule{0pt}{2.2ex}\hfil [\GeVns{}] \hfil & $\abs{y} < 0.3$ & $0.3 < \abs{y} < 0.6$ & $0.6 < \abs{y} < 0.9$ & $0.9 < \abs{y} < 1.2$\\ \hline
10--11 & 1.12E+03$\,\pm$0.3$\,\pm$7.9 & 1.06E+03$\,\pm$0.3$\,\pm$6.2 & 1.02E+03$\,\pm$0.3$\,\pm$4.6 & 8.84E+02$\,\pm$0.2$\,\pm$5.5 \\
11--12 & 6.55E+02$\,\pm$0.3$\,\pm$5.9 & 6.34E+02$\,\pm$0.3$\,\pm$4.8 & 6.20E+02$\,\pm$0.3$\,\pm$4.1 & 5.38E+02$\,\pm$0.3$\,\pm$4.7 \\
12--13 & 4.06E+02$\,\pm$0.3$\,\pm$5.0 & 3.97E+02$\,\pm$0.3$\,\pm$4.3 & 3.97E+02$\,\pm$0.3$\,\pm$3.9 & 3.39E+02$\,\pm$0.3$\,\pm$3.8 \\
13--14 & 2.65E+02$\,\pm$0.4$\,\pm$4.7 & 2.56E+02$\,\pm$0.4$\,\pm$4.1 & 2.54E+02$\,\pm$0.4$\,\pm$3.9 & 2.18E+02$\,\pm$0.4$\,\pm$3.8 \\
14--15 & 1.78E+02$\,\pm$0.4$\,\pm$4.5 & 1.71E+02$\,\pm$0.4$\,\pm$4.0 & 1.67E+02$\,\pm$0.4$\,\pm$3.9 & 1.46E+02$\,\pm$0.4$\,\pm$3.9 \\
15--16 & 1.21E+02$\,\pm$0.5$\,\pm$4.4 & 1.18E+02$\,\pm$0.5$\,\pm$3.9 & 1.14E+02$\,\pm$0.5$\,\pm$3.9 & 1.03E+02$\,\pm$0.5$\,\pm$3.3 \\
16--17 & 8.25E+01$\,\pm$0.6$\,\pm$4.4 & 8.19E+01$\,\pm$0.6$\,\pm$3.8 & 7.97E+01$\,\pm$0.6$\,\pm$3.9 & 7.00E+01$\,\pm$0.6$\,\pm$3.3 \\
17--18 & 6.05E+01$\,\pm$0.6$\,\pm$4.3 & 5.89E+01$\,\pm$0.6$\,\pm$3.8 & 5.76E+01$\,\pm$0.6$\,\pm$3.8 & 5.00E+01$\,\pm$0.7$\,\pm$3.3 \\
18--19 & 4.42E+01$\,\pm$0.7$\,\pm$4.3 & 4.30E+01$\,\pm$0.7$\,\pm$3.8 & 4.18E+01$\,\pm$0.7$\,\pm$3.8 & 3.64E+01$\,\pm$0.7$\,\pm$3.3 \\
19--20 & 3.25E+01$\,\pm$0.8$\,\pm$4.3 & 3.22E+01$\,\pm$0.8$\,\pm$3.8 & 3.11E+01$\,\pm$0.8$\,\pm$3.8 & 2.67E+01$\,\pm$0.9$\,\pm$3.3 \\
20--21 & 2.42E+01$\,\pm$0.9$\,\pm$4.3 & 2.46E+01$\,\pm$0.9$\,\pm$3.8 & 2.31E+01$\,\pm$0.9$\,\pm$3.8 & 2.03E+01$\,\pm$1.0$\,\pm$3.3 \\
21--22 & 1.92E+01$\,\pm$1.0$\,\pm$4.3 & 1.81E+01$\,\pm$1.0$\,\pm$3.8 & 1.80E+01$\,\pm$1.0$\,\pm$3.8 & 1.54E+01$\,\pm$1.1$\,\pm$3.3 \\
22--23 & 1.46E+01$\,\pm$1.2$\,\pm$4.3 & 1.40E+01$\,\pm$1.1$\,\pm$3.7 & 1.35E+01$\,\pm$1.2$\,\pm$3.8 & 1.20E+01$\,\pm$1.2$\,\pm$3.4 \\
23--24 & 1.12E+01$\,\pm$1.3$\,\pm$4.3 & 1.10E+01$\,\pm$1.3$\,\pm$3.7 & 1.07E+01$\,\pm$1.3$\,\pm$3.8 & 9.36E+00$\,\pm$1.4$\,\pm$3.4 \\
24--25 & 8.92E+00$\,\pm$1.4$\,\pm$4.4 & 8.75E+00$\,\pm$1.4$\,\pm$3.7 & 8.39E+00$\,\pm$1.4$\,\pm$3.8 & 7.46E+00$\,\pm$1.5$\,\pm$3.4 \\
25--26 & 7.43E+00$\,\pm$1.6$\,\pm$4.4 & 6.81E+00$\,\pm$1.6$\,\pm$3.7 & 6.86E+00$\,\pm$1.6$\,\pm$3.8 & 5.96E+00$\,\pm$1.7$\,\pm$3.4 \\
26--27 & 5.66E+00$\,\pm$1.8$\,\pm$4.4 & 5.45E+00$\,\pm$1.7$\,\pm$3.7 & 5.35E+00$\,\pm$1.8$\,\pm$3.8 & 4.96E+00$\,\pm$1.8$\,\pm$3.4 \\
27--28 & 4.72E+00$\,\pm$1.9$\,\pm$4.4 & 4.54E+00$\,\pm$1.9$\,\pm$3.7 & 4.26E+00$\,\pm$2.0$\,\pm$3.8 & 3.74E+00$\,\pm$2.1$\,\pm$3.4 \\
28--29 & 3.83E+00$\,\pm$2.1$\,\pm$4.4 & 3.70E+00$\,\pm$2.1$\,\pm$3.7 & 3.65E+00$\,\pm$2.1$\,\pm$3.8 & 3.08E+00$\,\pm$2.3$\,\pm$3.5 \\
29--30 & 3.04E+00$\,\pm$2.3$\,\pm$4.4 & 2.99E+00$\,\pm$2.3$\,\pm$3.7 & 2.91E+00$\,\pm$2.4$\,\pm$3.8 & 2.50E+00$\,\pm$2.5$\,\pm$3.5 \\
30--32 & 2.35E+00$\,\pm$1.9$\,\pm$4.4 & 2.35E+00$\,\pm$1.8$\,\pm$3.7 & 2.22E+00$\,\pm$1.9$\,\pm$3.9 & 1.90E+00$\,\pm$2.1$\,\pm$3.5 \\
32--34 & 1.69E+00$\,\pm$2.2$\,\pm$4.5 & 1.61E+00$\,\pm$2.2$\,\pm$3.7 & 1.53E+00$\,\pm$2.3$\,\pm$3.9 & 1.34E+00$\,\pm$2.4$\,\pm$3.5 \\
34--36 & 1.17E+00$\,\pm$2.6$\,\pm$4.5 & 1.19E+00$\,\pm$2.5$\,\pm$3.7 & 1.13E+00$\,\pm$2.6$\,\pm$3.9 & 9.62E-01$\,\pm$2.9$\,\pm$3.6 \\
36--38 & 8.70E-01$\,\pm$3.0$\,\pm$6.5 & 8.80E-01$\,\pm$2.9$\,\pm$5.9 & 8.32E-01$\,\pm$3.0$\,\pm$6.1 & 7.03E-01$\,\pm$3.4$\,\pm$5.8 \\
38--42 & 5.67E-01$\,\pm$2.6$\,\pm$6.5 & 5.51E-01$\,\pm$2.6$\,\pm$5.9 & 5.39E-01$\,\pm$2.6$\,\pm$6.1 & 4.70E-01$\,\pm$2.9$\,\pm$5.9 \\
42--46 & 3.34E-01$\,\pm$3.4$\,\pm$6.5 & 2.99E-01$\,\pm$3.5$\,\pm$5.9 & 3.13E-01$\,\pm$3.4$\,\pm$6.1 & 2.63E-01$\,\pm$3.7$\,\pm$5.9 \\
46--50 & 2.13E-01$\,\pm$4.4$\,\pm$6.5 & 1.87E-01$\,\pm$4.5$\,\pm$5.9 & 1.80E-01$\,\pm$6.5$\,\pm$6.1 & 1.64E-01$\,\pm$4.9$\,\pm$5.9 \\
50--60 & 1.00E-01$\,\pm$4.1$\,\pm$11 & 9.48E-02$\,\pm$4.1$\,\pm$11 & 8.37E-02$\,\pm$4.3$\,\pm$11 & 7.03E-02$\,\pm$4.9$\,\pm$11 \\
60--75 & 3.06E-02$\,\pm$6.4$\,\pm$11 & 2.97E-02$\,\pm$6.1$\,\pm$11 & 2.72E-02$\,\pm$6.5$\,\pm$11 & 2.39E-02$\,\pm$7.3$\,\pm$11 \\
75--95 & 7.00E-03$\,\pm$13$\,\pm$11 & 7.03E-03$\,\pm$12$\,\pm$11 & 9.80E-03$\,\pm$9.6$\,\pm$11 & 7.23E-03$\,\pm$12.0$\,\pm$11 \\
\end{scotch}
\end{table*}

\begin{table*}[htbp]
\centering
\topcaption{The \PsiTwo differential cross section times dimuon branching fraction
$\mathcal{B}\, \rd\sigma/\rd{}\pt$ in four rapidity ranges for the unpolarized scenario.
The relative uncertainties (first statistical and then systematic) are given in percent.
The systematic uncertainties are to be treated as bin-to-bin correlated.}
\label{tab:xs-PsiP-rap}
\begin{scotch}{ccccc}
\rule{0pt}{0.4cm}$\Delta \pt$ & \multicolumn{4}{c}{$\mathcal{B}\, \rd\sigma/\rd{}\pt$ [pb/\GeVns{}]} \\ \cline{2-5}
\rule{0pt}{2.2ex}\hfil [\GeVns{}] \hfil & $\abs{y} < 0.3$ & $0.3 < \abs{y} < 0.6$ & $0.6 < \abs{y} < 0.9$ & $0.9 < \abs{y} < 1.2$\\ \hline
10--11 & 4.07E+01$\,\pm$1.7$\,\pm$7.5 & 3.80E+01$\,\pm$1.7$\,\pm$6.2 & 3.82E+01$\,\pm$1.6$\,\pm$4.4 & 3.35E+01$\,\pm$1.5$\,\pm$4.5 \\
11--12 & 2.54E+01$\,\pm$1.6$\,\pm$5.8 & 2.48E+01$\,\pm$1.7$\,\pm$5.0 & 2.42E+01$\,\pm$1.7$\,\pm$4.1 & 2.13E+01$\,\pm$1.6$\,\pm$3.9 \\
12--13 & 1.62E+01$\,\pm$1.8$\,\pm$5.1 & 1.51E+01$\,\pm$1.9$\,\pm$4.6 & 1.58E+01$\,\pm$1.8$\,\pm$4.1 & 1.40E+01$\,\pm$1.8$\,\pm$3.8 \\
13--14 & 1.04E+01$\,\pm$2.0$\,\pm$4.8 & 1.07E+01$\,\pm$2.0$\,\pm$4.4 & 1.07E+01$\,\pm$2.0$\,\pm$4.0 & 8.88E+00$\,\pm$2.1$\,\pm$3.7 \\
14--15 & 7.77E+00$\,\pm$2.2$\,\pm$4.7 & 7.51E+00$\,\pm$2.2$\,\pm$4.3 & 6.98E+00$\,\pm$2.3$\,\pm$4.0 & 6.31E+00$\,\pm$2.4$\,\pm$3.6 \\
15--16 & 5.08E+00$\,\pm$2.6$\,\pm$4.8 & 4.97E+00$\,\pm$2.5$\,\pm$4.4 & 4.96E+00$\,\pm$2.6$\,\pm$4.3 & 4.13E+00$\,\pm$2.9$\,\pm$3.9 \\
16--17 & 3.79E+00$\,\pm$2.8$\,\pm$4.6 & 3.57E+00$\,\pm$2.9$\,\pm$4.3 & 3.42E+00$\,\pm$3.1$\,\pm$4.1 & 3.10E+00$\,\pm$3.2$\,\pm$3.7 \\
17--18 & 2.69E+00$\,\pm$3.2$\,\pm$4.7 & 2.63E+00$\,\pm$3.3$\,\pm$4.3 & 2.58E+00$\,\pm$3.4$\,\pm$4.3 & 2.16E+00$\,\pm$3.8$\,\pm$3.9 \\
18--19 & 1.94E+00$\,\pm$3.7$\,\pm$4.6 & 1.87E+00$\,\pm$3.8$\,\pm$4.2 & 1.96E+00$\,\pm$3.7$\,\pm$4.1 & 1.70E+00$\,\pm$4.1$\,\pm$3.7 \\
19--20 & 1.43E+00$\,\pm$4.3$\,\pm$4.7 & 1.30E+00$\,\pm$4.5$\,\pm$4.3 & 1.42E+00$\,\pm$4.3$\,\pm$4.2 & 1.23E+00$\,\pm$4.8$\,\pm$3.9 \\
20--22.5 & 9.07E-01$\,\pm$3.2$\,\pm$5.1 & 8.83E-01$\,\pm$3.3$\,\pm$4.7 & 8.96E-01$\,\pm$3.3$\,\pm$4.7 & 7.44E-01$\,\pm$3.9$\,\pm$4.3 \\
22.5--25 & 4.69E-01$\,\pm$4.4$\,\pm$5.2 & 5.05E-01$\,\pm$4.2$\,\pm$4.7 & 4.57E-01$\,\pm$4.5$\,\pm$4.7 & 4.08E-01$\,\pm$5.0$\,\pm$4.4 \\
25--27.5 & 2.81E-01$\,\pm$5.6$\,\pm$5.8 & 2.90E-01$\,\pm$5.4$\,\pm$5.4 & 2.75E-01$\,\pm$5.8$\,\pm$5.4 & 2.31E-01$\,\pm$6.8$\,\pm$5.1 \\
27.5--30 & 1.65E-01$\,\pm$7.2$\,\pm$5.7 & 1.66E-01$\,\pm$7.2$\,\pm$5.3 & 1.81E-01$\,\pm$7.1$\,\pm$5.3 & 1.44E-01$\,\pm$8.5$\,\pm$5.1 \\
30--35 & 8.83E-02$\,\pm$6.8$\,\pm$6.0 & 8.70E-02$\,\pm$7.2$\,\pm$5.5 & 8.40E-02$\,\pm$7.3$\,\pm$5.6 & 7.78E-02$\,\pm$8.0$\,\pm$5.4 \\
35--40 & 3.67E-02$\,\pm$10$\,\pm$7.8 & 2.95E-02$\,\pm$13$\,\pm$7.4 & 3.74E-02$\,\pm$11$\,\pm$7.5 & 3.50E-02$\,\pm$12$\,\pm$7.2 \\
40--55 & 9.96E-03$\,\pm$13$\,\pm$8.2 & 9.64E-03$\,\pm$13$\,\pm$7.9 & 1.03E-02$\,\pm$13$\,\pm$8.0 & 1.08E-02$\,\pm$14$\,\pm$7.8 \\
55--75 & 2.73E-03$\,\pm$22$\,\pm$12 & 2.14E-03$\,\pm$27$\,\pm$12 & 2.19E-03$\,\pm$28$\,\pm$12 & 3.29E-03$\,\pm$26$\,\pm$12 \\
\end{scotch}
\end{table*}

\begin{table*}[htbp]
\centering
\topcaption{The \PsiOne differential cross section times dimuon branching fraction
$\mathcal{B}\, \rd\sigma/\rd{}\pt$ for the integrated rapidity range $\abs{y} < 1.2$, in the unpolarized scenario.
The relative uncertainties (first statistical and then systematic) are given in percent.
The systematic uncertainties are to be treated as bin-to-bin correlated.
The average \pt values, $\langle \pt \rangle$, are calculated after acceptance and efficiency corrections.
Detector smearing has a negligible effect on this value.
The last three columns list the scaling factors needed to obtain the cross sections
corresponding to the polarization scenarios represented by the indicated
$\lambda_\vartheta^\mathrm{HX}$ values.}
\label{tab:xs-JPsi-rap0}
\begin{scotch}{*{6}{c}}
\rule{0pt}{0.4cm}$\Delta \pt$ & $\langle \pt \rangle$ & $\mathcal{B}\, \rd\sigma/\rd{}\pt$ & \multicolumn{3}{c}{$\lambda_\vartheta^\mathrm{HX}$ scaling factors} \\ \cline{4-6}
\rule{0pt}{2.2ex}\hfil [\GeVns{}] \hfil & [\GeVns{}] & [pb/\GeVns{}] & $+1$ & $-1$ & $0.10$\\ \hline
10--11 & 10.5 &  1.01E+03$\,\pm$0.1$\,\pm$7.9 & 1.31 & 0.68 & 1.03 \\
11--12 & 11.5 &  6.09E+02$\,\pm$0.1$\,\pm$5.9 & 1.30 & 0.68 & 1.03 \\
12--13 & 12.5 &  3.82E+02$\,\pm$0.2$\,\pm$5.0 & 1.29 & 0.69 & 1.03 \\
13--14 & 13.5 &  2.47E+02$\,\pm$0.2$\,\pm$4.7 & 1.28 & 0.70 & 1.03 \\
14--15 & 14.5 &  1.65E+02$\,\pm$0.2$\,\pm$4.5 & 1.26 & 0.71 & 1.03 \\
15--16 & 15.5 &  1.14E+02$\,\pm$0.2$\,\pm$4.4 & 1.25 & 0.71 & 1.03 \\
16--17 & 16.5 &  7.84E+01$\,\pm$0.3$\,\pm$4.4 & 1.24 & 0.72 & 1.03 \\
17--18 & 17.5 &  5.66E+01$\,\pm$0.3$\,\pm$4.3 & 1.23 & 0.73 & 1.02 \\
18--19 & 18.5 &  4.13E+01$\,\pm$0.4$\,\pm$4.3 & 1.22 & 0.73 & 1.02 \\
19--20 & 19.5 &  3.05E+01$\,\pm$0.4$\,\pm$4.3 & 1.21 & 0.74 & 1.02 \\
20--21 & 20.5 &  2.30E+01$\,\pm$0.5$\,\pm$4.3 & 1.20 & 0.75 & 1.02 \\
21--22 & 21.5 &  1.76E+01$\,\pm$0.5$\,\pm$4.3 & 1.19 & 0.75 & 1.02 \\
22--23 & 22.5 &  1.35E+01$\,\pm$0.6$\,\pm$4.3 & 1.19 & 0.76 & 1.02 \\
23--24 & 23.5 &  1.05E+01$\,\pm$0.6$\,\pm$4.3 & 1.18 & 0.77 & 1.02 \\
24--25 & 24.5 &  8.35E+00$\,\pm$0.7$\,\pm$4.4 & 1.17 & 0.77 & 1.02 \\
25--26 & 25.5 &  6.75E+00$\,\pm$0.8$\,\pm$4.4 & 1.17 & 0.78 & 1.02 \\
26--27 & 26.5 &  5.35E+00$\,\pm$0.9$\,\pm$4.4 & 1.16 & 0.78 & 1.02 \\
27--28 & 27.5 &  4.31E+00$\,\pm$1.0$\,\pm$4.4 & 1.16 & 0.79 & 1.02 \\
28--29 & 28.5 &  3.57E+00$\,\pm$1.1$\,\pm$4.4 & 1.15 & 0.79 & 1.02 \\
29--30 & 29.5 &  2.86E+00$\,\pm$1.2$\,\pm$4.4 & 1.15 & 0.80 & 1.02 \\
30--32 & 30.9 &  2.21E+00$\,\pm$0.9$\,\pm$4.4 & 1.14 & 0.80 & 1.02 \\
32--34 & 32.9 &  1.55E+00$\,\pm$1.1$\,\pm$4.5 & 1.13 & 0.81 & 1.02 \\
34--36 & 35.0 &  1.11E+00$\,\pm$1.3$\,\pm$4.5 & 1.12 & 0.82 & 1.01 \\
36--38 & 37.0 &  8.22E-01$\,\pm$1.5$\,\pm$6.5 & 1.12 & 0.83 & 1.01 \\
38--42 & 39.8 &  5.33E-01$\,\pm$1.3$\,\pm$6.5 & 1.11 & 0.83 & 1.01 \\
42--46 & 43.8 &  3.02E-01$\,\pm$1.8$\,\pm$6.5 & 1.10 & 0.85 & 1.01 \\
46--50 & 47.9 &  1.86E-01$\,\pm$2.3$\,\pm$6.5 & 1.09 & 0.86 & 1.01 \\
50--60 & 54.2 &  8.75E-02$\,\pm$2.1$\,\pm$10.9 & 1.08 & 0.87 & 1.01 \\
60--75 & 66.0 &  2.78E-02$\,\pm$3.2$\,\pm$11.1 & 1.07 & 0.89 & 1.01 \\
75--95 & 82.9 &  7.97E-03$\,\pm$5.4$\,\pm$11.2 & 1.05 & 0.91 & 1.01 \\
95--120 & 104.1 &  1.96E-03$\,\pm$10.7$\,\pm$11.4 & 1.04 & 0.92 & 1.01 \\
\end{scotch}
\end{table*}

\begin{table*}[ht!]
\centering
\topcaption{The \PsiTwo differential cross section times dimuon branching fraction
$\mathcal{B}\, \rd\sigma/\rd{}\pt$ for the integrated rapidity range $\abs{y} < 1.2$, in the unpolarized scenario.
The relative uncertainties (first statistical and then systematic) are given in percent.
The systematic uncertainties are to be treated as bin-to-bin correlated.
The average \pt values, $\langle \pt \rangle$, are calculated after acceptance and efficiency corrections.
Detector smearing has a negligible effect on this value.
The last three columns list the scaling factors needed to obtain the cross sections
corresponding to the polarization scenarios represented by the indicated
$\lambda_\vartheta^\mathrm{HX}$ values.}
\label{tab:xs-PsiP-rap0}
\begin{scotch}{*{6}{c}}
\rule{0pt}{0.4cm}$\Delta \pt$ & $\langle \pt \rangle$ & $\mathcal{B}\, \rd\sigma/\rd{}\pt$ & \multicolumn{3}{c}{$\lambda_\vartheta^\mathrm{HX}$ scaling factors} \\ \cline{4-6}
\rule{0pt}{2.2ex}\hfil [\GeVns{}] \hfil & [\GeVns{}] & [pb/\GeVns{}] & $+1$ & $-1$ & $0.03$\\ \hline
10--11 & 10.5 &  3.80E+01$\,\pm$0.8$\,\pm$7.5 & 1.31 & 0.68 & 1.01 \\
11--12 & 11.5 &  2.41E+01$\,\pm$0.8$\,\pm$5.8 & 1.30 & 0.69 & 1.01 \\
12--13 & 12.5 &  1.54E+01$\,\pm$0.9$\,\pm$5.1 & 1.28 & 0.69 & 1.01 \\
13--14 & 13.5 &  1.02E+01$\,\pm$1.0$\,\pm$4.8 & 1.27 & 0.70 & 1.01 \\
14--15 & 14.5 &  7.15E+00$\,\pm$1.1$\,\pm$4.7 & 1.26 & 0.71 & 1.01 \\
15--16 & 15.5 &  4.79E+00$\,\pm$1.3$\,\pm$4.8 & 1.25 & 0.72 & 1.01 \\
16--17 & 16.5 &  3.48E+00$\,\pm$1.5$\,\pm$4.6 & 1.24 & 0.72 & 1.01 \\
17--18 & 17.5 &  2.52E+00$\,\pm$1.7$\,\pm$4.7 & 1.23 & 0.73 & 1.01 \\
18--19 & 18.5 &  1.87E+00$\,\pm$1.9$\,\pm$4.6 & 1.22 & 0.74 & 1.01 \\
19--20 & 19.5 &  1.34E+00$\,\pm$2.2$\,\pm$4.7 & 1.21 & 0.74 & 1.01 \\
20--22.5 & 21.1 &  8.57E-01$\,\pm$1.7$\,\pm$5.1 & 1.20 & 0.75 & 1.01 \\
22.5--25 & 23.6 &  4.61E-01$\,\pm$2.2$\,\pm$5.2 & 1.18 & 0.77 & 1.01 \\
25--27.5 & 26.1 &  2.69E-01$\,\pm$2.9$\,\pm$5.8 & 1.16 & 0.78 & 1.01 \\
27.5--30 & 28.7 &  1.65E-01$\,\pm$3.7$\,\pm$5.7 & 1.15 & 0.79 & 1.01 \\
30--35 & 32.2 &  8.42E-02$\,\pm$3.6$\,\pm$6.0 & 1.13 & 0.81 & 1.00 \\
35--40 & 37.2 &  3.47E-02$\,\pm$5.8$\,\pm$7.8 & 1.12 & 0.83 & 1.00 \\
40--55 & 45.5 &  1.02E-02$\,\pm$6.6$\,\pm$8.2 & 1.10 & 0.85 & 1.00 \\
55--75 & 62.4 &  2.35E-03$\,\pm$12.7$\,\pm$12.3 & 1.07 & 0.88 & 1.00 \\
75--100 & 84.1 &  5.62E-04$\,\pm$24.4$\,\pm$12.6 & 1.05 & 0.91 & 1.00 \\
\end{scotch}
\end{table*}

\begin{table*}[htbp]
\centering
\topcaption{The ratio of the \PsiTwo to \PsiOne differential cross sections times dimuon branching fractions
in percent, as a function of \pt,
in the unpolarized scenario for $\abs{y} < 1.2$.
The first uncertainty is statistical and the second is systematic.
The systematic uncertainties are to be treated as bin-to-bin correlated.}
\label{tab:ratio-xs}
\begin{scotch}{ccc}
\rule{0pt}{0.4cm}$\Delta \pt$ & $\langle \pt \rangle$&
$[\mathcal{B}\, \sigma(\PsiTwo)] / [\mathcal{B}\,\sigma(\PsiOne)]$ \\
\hfil [\GeVns{}] \hfil & [\GeVns{}] & [\%] \\ \hline
10--11 & 10.5 &  3.75$\,\pm$0.03$\,\pm$0.11 \\
11--12 & 11.5 &  3.93$\,\pm$0.03$\,\pm$0.11 \\
12--13 & 12.5 &  4.04$\,\pm$0.04$\,\pm$0.11 \\
13--14 & 13.5 &  4.11$\,\pm$0.04$\,\pm$0.11 \\
14--15 & 14.5 &  4.30$\,\pm$0.05$\,\pm$0.12 \\
15--16 & 15.5 &  4.20$\,\pm$0.06$\,\pm$0.11 \\
16--17 & 16.5 &  4.39$\,\pm$0.07$\,\pm$0.12 \\
17--18 & 17.5 &  4.42$\,\pm$0.08$\,\pm$0.12 \\
18--19 & 18.5 &  4.45$\,\pm$0.09$\,\pm$0.12 \\
19--20 & 19.5 &  4.37$\,\pm$0.10$\,\pm$0.11 \\
20--22.5 & 21.1 &4.49$\,\pm$0.08$\,\pm$0.05 \\
22.5--25 & 23.6 &4.58$\,\pm$0.10$\,\pm$0.05 \\
25--27.5 & 26.1 &4.69$\,\pm$0.14$\,\pm$0.04 \\
27.5--30 & 28.7 &4.85$\,\pm$0.18$\,\pm$0.05 \\
30--35 & 32.2 &  4.84$\,\pm$0.18$\,\pm$0.05 \\
35--40 & 37.2 &  4.47$\,\pm$0.26$\,\pm$0.05 \\
40--55 & 45.5 &  4.47$\,\pm$0.30$\,\pm$0.04 \\
55--75 & 62.3 &  6.08$\,\pm$0.80$\,\pm$0.12 \\
75--100 & 82.9 & 7.64$\,\pm$1.98$\,\pm$0.41 \\
\end{scotch}
\end{table*}

    } \cleardoublepage \section{The CMS Collaboration \label{app:collab}}\begin{sloppypar}\hyphenpenalty=5000\widowpenalty=500\clubpenalty=5000\textbf{Yerevan Physics Institute,  Yerevan,  Armenia}\\*[0pt]
V.~Khachatryan, A.M.~Sirunyan, A.~Tumasyan
\vskip\cmsinstskip
\textbf{Institut f\"{u}r Hochenergiephysik der OeAW,  Wien,  Austria}\\*[0pt]
W.~Adam, T.~Bergauer, M.~Dragicevic, J.~Er\"{o}, M.~Friedl, R.~Fr\"{u}hwirth\cmsAuthorMark{1}, V.M.~Ghete, C.~Hartl, N.~H\"{o}rmann, J.~Hrubec, M.~Jeitler\cmsAuthorMark{1}, W.~Kiesenhofer, V.~Kn\"{u}nz, M.~Krammer\cmsAuthorMark{1}, I.~Kr\"{a}tschmer, D.~Liko, I.~Mikulec, D.~Rabady\cmsAuthorMark{2}, B.~Rahbaran, H.~Rohringer, R.~Sch\"{o}fbeck, J.~Strauss, W.~Treberer-Treberspurg, W.~Waltenberger, C.-E.~Wulz\cmsAuthorMark{1}
% ...
\vskip\cmsinstskip
% ...
1:~~Also at Vienna University of Technology, Vienna, Austria\\
2:~~Also at CERN, European Organization for Nuclear Research, Geneva, Switzerland\\
% ...

\end{sloppypar}
\end{document}